\documentclass[preprintnumbers,twocolumn,superscriptaddress,showpacs,amsmath]{revtex4}
\usepackage{graphicx}
\usepackage{dcolumn}
\usepackage{bm}

\usepackage{graphicx}
\usepackage{epstopdf}
\usepackage{dcolumn}
\usepackage{bm}
\usepackage{subfigure}
\usepackage{color} 

\begin{document}
\title{Exact analysis of the spectral properties of  the  anisotropic two-bosons Rabi model}

\author{Shuai Cui}
\affiliation{Beijing National Laboratory for Condensed Matter Physics,\\
and Institute of Physics, Chinese Academy of Sciences, Beijing
100190, China}
\author{Jun-Peng Cao}
\affiliation{Beijing National Laboratory for Condensed Matter Physics,\\
and Institute of Physics, Chinese Academy of Sciences, Beijing
100190, China} \affiliation{Collaborative Innovation Center of
Quantum Matter, Beijing 100190, China}

\author{Heng Fan}
\email{hfan@iphy.ac.cn}
\affiliation{Beijing National Laboratory for Condensed Matter Physics,\\
and Institute of Physics, Chinese Academy of Sciences, Beijing
100190, China} \affiliation{Collaborative Innovation Center of
Quantum Matter, Beijing 100190, China}

\author{Luigi Amico}
\email{lamico@dmfci.unict.it} \affiliation{CNR-MATIS-IMM \&
Dipartimento di Fisica e Astronomia, Universit\'a Catania, Via S.
Sofia 64, 95127 Catania, Italy} 
\affiliation{INFN-Laboratori Nazionali del Sud, INFN, via S. Sofia 62, 95123 Catania, Italy}
\affiliation{Center for Quantum
Technologies, National University of Singapore, 3 Science Drive 2,
117543 Singapore} 

\date{\today}

\begin{abstract}
We introduce the anisotropic two-photon Rabi model in which the rotating and counter rotating terms enters along with two different coupling constants.
Eigenvalues and eigenvectors are studied with exact means. We employ a variation of the Braak method based on Bogolubov rotation of the underlying $su(1,1)$ Lie algebra.
Accordingly, the spectrum is provided by the analytical properties of a suitable meromorphic function.    Our formalism applies to the two-modes Rabi model as well, sharing the same algebraic structure of the two-photon model.
Through the analysis of the spectrum, we discover that the model displays close analogies to many-body systems undergoing quantum phase transitions. 

\end{abstract}

 \maketitle

\section{introduction}

Two-photon Rabi type models serve the study of various quantum effects in systems of  bosonic fields coupled to a  set
of localized levels.  Its simplest instance traces back to the theory of micromaser \cite{Scully,HarochePRL}.
In such a  quantum optics set up,  three atomic levels are  coupled each other
through two-boson  field in a cascade transition $|g\rangle \rightarrow |i\rangle
\rightarrow |e\rangle$, with the energy difference $\omega_{eg}$ set to
twice of boson field frequency $\omega$, and with the intermediate state being
strongly detuned with $\omega_{ei}$ and $\omega_{ig}$. After
adiabatically eliminating the intermediate state, and neglecting the Stark shift,  one arrives at the
two-photon Rabi model \cite{Ashraf1990}. Phenomenologically, such type of  models describe two-level atom interacting  with squeezed light\cite{gerry}.
More recently, models of these type emerged in different context of quantum technology. Indeed, they can be obtained as  effective models describing quantum dots inserted in QED micro cavity \cite{morigi};
there, the localized levels are provided by the excitonic energies of the quantum dot; two photon process in quantum dots were  observed through photoluminiscence experiments \cite{two-photon_experiments}.
The  two-mode Rabi models also emerge in circuit QED involving superconducting qubits in the ultra strong regime
where non-linear couplings become realistic \cite{Mooji,Niemczyk,Grimsmo} and in ion traps \cite{twophoton_ions}.
Finally, the two-mode boson field can
be realized by a charged particle in a magnetic field \cite{Novaes}.

In a weak coupling limit between the bosonic degree of freedom and the localized levels (that can be modelled by a  spin degree of freedom), the features of the system can be obtained  employing the so-called Rotating-Wave Approximation (RWA)\cite{rwa}.
In most, if not all, of the scenarios depicted above, however,
the `physical working point' is in a strong coupling regime,
where the RWA is a poor approximation. The limits of the RWA for the two-boson Rabi models were investigated numerically in \cite{Ng}.

Here, we deal  with the exact energy spectrum and eigenstates of the two-photon Rabi model.
The exact solvability beyond RWA was investigated by means of Algebraic Bethe Ansatz in \cite{ABA}.
 More recently, the exact eigenenergies and eigenstates were obtained  in the `isotropic' case where the rotating terms and counter-rotating terms have the same weight \cite{Travenec,twoph_isotropic_exact,ZhangYZ}.
The solution is obtained applying a variation of the procedure that  Braak recently devised for the single photon Rabi model \cite{Braak} (see also \cite{Moroz,Chen}).

In this paper, we introduce a class of anisotropic version of the two-mode
and two-photon Rabi model, where the rotating and counter rotating terms play along with two different parameters. As the single-mode case,  our model enjoys a $Z_2$ parity symmetry.  Relying on it, we will give the exact solution of anisotropic
two-photon and two-mode Rabi model by Bogolubov operator method \cite{Chen,Emary}.  We will demonstrate that such a model the energy levels changes a function of a suitable control parameter with  a  similar effects displayed in quantum phase transitions of many-body systems. A discontinuous transition is controlled by the anisotropic parameter.
For higher values of the spin-boson strength,  we find a condensation of levels  breaking  implying a  super-radiance phenomenon.

The paper is organized as follows. In Sec. II, the mathematical implications arising from the relation between the  $su(1,1)$ Lie algebra and the two-photon Rabi model is discussed. In Sec. III, we
present our approach in detail for the anisotropic two-photon  Rabi
model. In Sec. IV, we discuss the kind of 'critical behavior' that we found in the energy spectra of the model. In Sec. V, we draw our conclusions and future directions. In the Appendix A we discuss a specific  circuit QED realization of the model Hamiltonian.   In Appendix B, we sketch the application of our approach to the two-mode Rabi model.
\section{ The two-photon  Rabi model: $su(1,1)$ and $Z_2$ symmetry}
\label{sect2}
The Hamiltonian reads
\begin{eqnarray}
H_{2-ph}&=&  \omega a^\dagger a+\Delta \sigma_z +g(\sigma^+ a^2
+\sigma^- a^{\dagger 2}) \nonumber \\
&& \qquad \qquad +\lambda g (\sigma^+ a^{\dagger 2} +\sigma^- a^2),
\label{H2photon}
\end{eqnarray}
where the  $\omega$ fix  the energy of the  bosonic  field  and  $g$ is the coupling
constant, $\lambda$ being the anisotropic parameter characterizing the difference between rotating and counter-rotating spin-boson interactions.

The features of the Hamiltonian above are intimately connected with $su(1,1)$ Lie algbra realized by
\begin{eqnarray}
K_0=\frac{1}{2}(a^\dagger a+\frac{1}{2}), \  K_+=\frac{1}{2}
a^{\dagger 2}, \  K_-=\frac{1}{2} a^2,
\end{eqnarray}
obeying to
\begin{eqnarray}
[K_0, K_\pm]=\pm K_\pm, \qquad [K_+, K_-]=-2K_0,
\label{su11_comm}
\end{eqnarray}
with invariant Casimir operator
\begin{eqnarray}
{\cal C}=K_+ K_- +K_0 (1-K_0)=\kappa (1-\kappa),
\end{eqnarray}
where $\kappa$ labels the representations.  $K_\pm$ and  $K_0$ are off-diagonal and diagonal operators in the Cartan basis, respectively. Relations (\ref{su11_comm}) can be realized through first order differential  operators acting on a suitable functional space: \begin{equation}
K_0 \leftrightarrow  z \frac{d}{dz} +\kappa\, , \, K_+\leftrightarrow  z^2  \frac{d}{dz} +2 \kappa z \, , \, K_- \leftrightarrow    \frac{d}{dz} \, .
\label{su11_bargman}
\end{equation}
The isomorphism above  can be seen as a `Bargman realization' for $su(1,1)$ (see for example \cite{Wybourne}).
We remark that $su(1,1)$ is a non-compact algebra and therefore the representation space is a  non trivial manifold.
In the $su(1,1)$ specific case, the representation space is isomorphic to an hyperboloid. Different representations correspond to a discrete series in the upper or lower branch of the hyperboloid, often denoted as $D^\pm_k$. In the case of the two-boson case considered here, $\kappa=\frac{1}{4}, \frac{3}{4}$ for  even or  odd number Fock basis,
respectively (see the appendix for the representation index for the su(1,1) Rabi model realized by two-bosonic-mode).
Another possible, but inequivalent,  representation is  the  continuous one, labeled by the eigenvalues of $K_\pm$ \cite{Limblad-Nagel}).
As we shall see in the next sections, the computation of the exact solution of the model (\ref{H2mode}) involves the diagonalization inside the $su(1,1)$ algebra.
We will choose a representation in which   $K_0$ is diagonal. Correspondingly, certain constraints will emerge in the solution (see (\ref{condition})).  In appendix, we  discuss the spectrum  beyond such constraint.

The Hamiltonian (\ref{H2photon} )  enjoys a $Z_2$ symmetry:
\begin{eqnarray}
\label{twoph_Z2}
\Pi_{2-ph}&=&e^{i\frac{\pi}{2}(a^\dagger a +\sigma_z+1)} \nonumber \\
&=&-\sigma_z
\left[\cos (\frac{\pi}{2} a^\dagger a)+i\sin (\frac{\pi}{2}
a^\dagger a)\right],
\end{eqnarray}
whose eigenvalues belong to a four-dimensional manifold spanned by $p=\pm 1$ and $\pm i$. Such a manifold yields four
irreducible subspaces of the $su(1,1)$ for two photons. Incidentally, we note that  the model with $Z_2$ symmetry breaking, i.e. with  $\sigma_x$ term added to  the Hamiltonian, 
can also be solved by our method \cite{Xie}.

The following observation is important  for the procedure  to analyze the eigen-system of (\ref{H2photon}).
Despite, the (unitary) transformation
\begin{equation}
\label{Z2trasf_twomode}
a^\dagger \rightarrow i a^\dagger \quad , \quad
a\rightarrow -i a\quad 
\end{equation}
 is not a symmetry of the Hamiltonian, it defines an automorphism of $su(1,1)$. In this way, we shall see that (\ref{Z2trasf_twomode}) induces a specific redundancy of the eigevectors  of  (\ref{H2photon}), that is ultimately
 useful for us.
 The issue is more evident by spelling-out the spin  basis in (\ref{H2photon})
 \begin{equation}
 \label{H2modeW}
(H)=\left(\begin{array}{cc} \omega a^\dagger a +\Delta & g (a^2  +\lambda a^{\dagger 2} )\\
 g (a^{\dagger 2}  +\lambda a^{2}) &  \omega a^\dagger a -\Delta \end{array} \right)
\end{equation}
and applying a rotation of the basis: $ \displaystyle{
W=\frac{1}{\sqrt{2}}\left(\begin{array}{cc} 1& -1 \\ 1 &
1
\end{array} \right) }$:
\begin{widetext}
\begin{equation}
W^\dagger (H) W= \frac{1}{2} \left(\begin{array}{cc} 2\omega a^\dagger a +g(1+\lambda) (a^2 +a^{\dagger 2})&-2 \Delta- g (\lambda -1)(a^2  -a^{\dagger 2} )\\
-2 \Delta + g (\lambda -1)(a^2  -a^{\dagger 2} )\ &  2\omega a^\dagger a -g(1+\lambda) (a^2 +a^{\dagger 2}) \end{array} \right)
 \end{equation}
\end{widetext}

By inspection of the equation above, we see that its  eigen-system
\begin{equation}
W^\dagger (H) W  \left(\begin{array}{c} v'_1 \\ v'_2 \end{array} \right)= E \left(\begin{array}{c} v'_1 \\ v'_2 \end{array} \right)
\end{equation}
is left invariant   by the transformation (\ref{Z2trasf_twomode})  by simultaneously swapping $v_1$ and  $v_2$.

Summarizing: the $su(1,1)$ automorphism (\ref{Z2trasf_twomode}) implies the existence of two sets of eigenstates of (\ref{H2photon}): the original ones  $\displaystyle{\left (v_1, v_2\right )^T}$  and $\displaystyle{ \left (v'_2 , v'_1\right )^T}$ where
$\displaystyle{\left (v'_1, v'_2  \right )^T =W^\dagger \left (v_1,  v_2\right )^T}$.

Equivalently,  the role of (\ref{Z2trasf_twomode}) in the Bogolubov scheme we will adopt below  is played by the transformation $z\rightarrow -z$ in the  $su(1,1)$-Bargman realization (\ref{su11_bargman}) of the spectral problem for  (\ref{H2modeW}).
In this way, the redundancy property of the eigenvectors involves $\displaystyle{\left (v_1 (z), v_2(z)\right )^T}$  and $\displaystyle{ \left (v'_2(-z) , v'_1(-z)\right )^T}$.
Indeed, the resulting   structure  of the Hilbert space implies that the domain of analyticity of the eigenvectors $\left (v_1 (z), v_2(z)\right )^T$ can be extended to the whole complex plane.

In the next section, we shall see how such a property will be exploited to construct a meromorphic function ($G(z)$) whose analytical structure provides the spectrum of the Hamiltonian (\ref{H2photon}).

\section{Exact analysis of the eigensystem}
In this section, we will provide the exact analysis of spectrum $\&$ eigenstates  of  the anisotropic two-mode Rabi model  Eq.(\ref{H2photon}). We will be  formulating an ansatz for the eigenvectors
expressed as a series expansion  that we eventually determine through recurrence relations.
It turns out that such recurrence relations are easier to solve if we rotate the spin axes  suitably:  $(H)'=U(H)U^{\dagger}$, with
$\displaystyle{
U=\left(\begin{array}{cc} \cos\beta & -\sin\beta \\ \sin\beta &
\cos\beta
\end{array} \right) }
$: 
\begin{widetext}
\begin{eqnarray}
\label{2photonH2}
(H)'=
\left(\begin{array}{cc} \omega a^\dagger a +p +r(a^2 +a^{\dagger 2})
& -q +s a^2 +t a^{\dagger 2}
\\ -q +s
a^{\dagger 2} +t a^{2} & \omega a^\dagger a -p -r(a^2 +a^{\dagger
2})
\end{array} \right)
\end{eqnarray}
where
$p =\Delta \cos 2\beta, \ q= \Delta \sin 2\beta$,
$\displaystyle{r =\frac{\sin 2\beta}{2} (1+\lambda)g}$,
$s=(\cos^2 \beta -\lambda \sin^2 \beta) g$, and
$t =(\lambda \cos^2 \beta -\sin^2 \beta) g$.
\end{widetext}
Next, we perform a rotation in the $su(1,1)$ algebra by using a Bogolubov transformation\cite{Xie} ,
\begin{eqnarray}
\alpha=ua+va^\dagger, \qquad \alpha^\dagger=ua^\dagger +va,
\end{eqnarray}
where $\left[ \alpha,\alpha^\dagger\right ] =1$ if $u^2-v^2=1$. Because of the anisotropy, we notice  that we cannot get rid of  the non-Cartan generators of $su(1,1)$ in all the matrix elements.  Nevertheless, we can fix the parameters $u,v$, and $\beta$ in such a way that the upper (lower) off-diagonal  element of $(H)'$  is a  generalized lowering (raising) operator in  $su(1,1)$.
Such a choice will simplify the solution of the recurrence relations below (see (\ref{2photonEq1}), (\ref{2photonEq2}), (\ref{2photonK_m})).  The Hamiltonian, then, reads
\begin{widetext}
\begin{eqnarray} 
(H)'= \left(\begin{array}{cc}\omega\eta(\alpha^\dagger \alpha +\frac{1}{2})-\frac{\omega}{2}+\Delta \cos 2\beta & -\Delta \sin 2\beta -r(1-\lambda)\frac{g}{\omega}(2\alpha^\dagger \alpha+1)+(1-\lambda) g \alpha^2, \\
-\Delta \sin 2\beta-r(1-\lambda)\frac{g}{\omega}(2\alpha^\dagger \alpha+1)+(1-\lambda) g \alpha^{\dagger 2}& \omega(\frac{2}{\eta}-\eta) (\alpha^\dagger
\alpha+\frac{1}{2})-\frac{\omega}{2} -\Delta \cos 2\beta
-\frac{2r}{\eta}(\alpha^{\dagger 2}+\alpha^2)\end{array} \right) \nonumber
\end{eqnarray}
\end{widetext}
with
\begin{eqnarray}
\label{u}
u=\sqrt{\frac{1+\eta}{2\eta}}, \ v=\sqrt{\frac{1-\eta}{2\eta}},
\ \cos 2\beta=\frac{1-\lambda}{1+\lambda}\eta,
\end{eqnarray}
where
\begin{equation}
\label{z}
\eta=\sqrt{\frac{1-(1+\lambda)^2 g^2/\omega^2}{1-(1-\lambda)^2
g^2/\omega^2}},
\end{equation}
with the condition
\begin{eqnarray}
\label{condition}
|g|< \frac{\omega}{|1+\lambda|}\,.
\end{eqnarray}
As announced in Sect.\ref{sect2}, the constraint (\ref{condition}) arises because of the non trivial topology of the $su(1,1)$ representation space.

The Fock basis for the new bosonic operators $\alpha$ is spanned by
$
|m\rangle_\alpha = (\alpha^\dagger)^m |0\rangle_\alpha \; ,\;
m=0,1,2,...,
$
where the vacuum
$\alpha |0\rangle_\alpha=0$ is
\begin{eqnarray}\label{psi}
|0\rangle_\alpha =\frac{1}{\sqrt{u}} \sum_{n=0}^\infty
(-\frac{v}{u})^n
\frac{\sqrt{(2n)!}}{2^n n!}|2 n\rangle\; . \label{psi0}
\end{eqnarray}
We observe that, although the   normalization coefficient can be  fixed by the series expansion of $1/\sqrt{u}$,  
we   work with non-normalized Fock basis (the normalization constant does not change  the results).

We are now ready to formulate the ansatz for the eigenvectors of $(H)'$:
\begin{equation}
(H)'\left(\begin{array}{c} \phi_1 \\ \phi_2 \end{array} \right)=E\left(\begin{array}{c} \phi_1 \\ \phi_2 \end{array} \right) \; ,
\end{equation}
\begin{eqnarray}\label{su11_lamphi}
\left(\begin{array}{c} \phi_1 \\ \phi_2 \end{array} \right)
=\left(\begin{array}{c} \sum_{m=0}^\infty L_m |m\rangle_\alpha\\
\
\\ \sum_{m=0}^\infty K_m |m\rangle_\alpha \end{array}
\right),
\end{eqnarray}
where $K_m$ and $L_m$ are coefficients to be determined.

Plugging  (\ref{su11_lamphi}) into the spectral problem of the Hamiltonian
(\ref{2photonH2}), we obtain,
\begin{widetext}
\begin{eqnarray}
&&\sum_{m=0} f_mL_m |m\rangle_\alpha -\sum_{m=0}d_mK_m|m\rangle_\alpha +(1-\lambda)g\sum_{m=0} m(m-1)K_m|m-2\rangle_\alpha
=0, \label{2photonEq1} \\
&&-\sum_{m=0}d_mL_m
|m\rangle_\alpha +(1-\lambda)g\sum_{m=0} L_m|m+2\rangle_\alpha  +  \label{2photonEq2} \\
&&\hspace*{2cm}\sum_{m=0}\left[\omega\left(\frac{2}{\eta}-\eta \right)(m+\frac{1}{2})-\frac{\omega}{2} -p -E \right]K_m |m\rangle_\alpha
-\frac{2r}{\eta}\sum_{m=0}K_m \left[|m+2\rangle_\alpha
+m(m-1)|m-2\rangle_\alpha \right]=0\, , \nonumber
\end{eqnarray}
where
$\displaystyle{
d_m= q+r(1-\lambda)(2m+1)\frac{g}{\omega}\; ,
f_m=\omega\eta(m+\frac{1}{2})-\frac{\omega}{2}+p-E}
$
and  $[\alpha, \alpha^{\dagger m}]=m
\alpha^{\dagger m-1}$ were exploited.
\end{widetext}
By inspection of Eq. (\ref{2photonEq1}), we obtain
\begin{eqnarray}
L_m = \frac{d_mK_m -(1-\lambda)(m+2)(m+1)gK_{m+2}}{f_m},
\end{eqnarray}
leading to a closed Eq.(\ref{2photonEq2}) for $K_m$:
\begin{eqnarray} \label{2photonK_m}
 a_m K_{m+2} &=& b_m K_m +c_m K_{m-2},
 \end{eqnarray}
 \begin{eqnarray}
 a_m &=&\left[-\frac{d_m(1-\lambda)g}{f_m} +\frac{2r}{\eta}
\right](m+2)(m+1), \nonumber \\
 b_m &=&-\frac{d_m^2}{f_m} -\frac{(1-\lambda)^2 m(m-1)g^2}{f_{m-2}}
\nonumber \\ && \qquad +\omega \left(\frac{2}{\eta}
-\eta\right)\left(m+\frac{1}{2}\right) -\frac{\omega}{2} -p -E, \nonumber \\
c_m &=& \frac{(1-\lambda)g d_{m-2}}{f_{m-2}} -\frac{2r}{\eta}.
\nonumber
\end{eqnarray}
With a similar logic employed in \cite{Braak,Chen,Xie}, we want to extract the eigenvalues of $H$ by looking at the analytical structure of a meromorphic function $G(z)$,  constructed by imposing that  the eigenvectors of the Hamiltonian are indeed analytic in whole complex plane.
In our scheme, $G(z)$  is constructed resorting the property of the eigen-system under the application of the $su(1,1)$ automorphism (\ref{Z2trasf_twomode}).
Namely, the transformation (\ref{Z2trasf_twomode})  on the new bosonic operators $\alpha, \alpha^\dagger$ leaves the Hamiltonian's  spectrum invariant, with  eigenvectors changing  as
\begin{eqnarray}
\left(\begin{array}{c}\overline{ \varphi}_1 \\ \overline{\varphi}_2 \end{array} \right) = C \left(\begin{array}{c} \varphi_2 \\ \varphi_1 \end{array} \right) \, ,\quad
\left(\begin{array}{c} \varphi_1 \\ \varphi_2 \end{array} \right)
=W^\dagger U \left(\begin{array}{c} \phi_1 \\ \phi_2 \end{array}
\right) \, .
\end{eqnarray}
Resorting to the parity symmetry operator $\Pi_{2-ph}$, in the even Fock space of
$\alpha$, the constant  $C=\pm 1$, while  it is $C=\pm i$,  for  Fock
states with odd parity.
The transcendental functions in even and odd parity sectors are
\begin{eqnarray}
\label{G-functions}
&&G_{\lambda,\pm}^e =\langle 0|[\overline{\varphi}_2 \mp \varphi_1 ], \\
&&G_{\lambda,\pm}^o =\langle 1|[-i\overline{\varphi}_2 \mp \varphi_1
],
\end{eqnarray}
where $G_{\lambda,\pm}^o$ is defined real by multiplying
an unimportant factor $-i$. Indeed, even and odd functions
can be considered on equal footing multiplying by the overall factor $D_m$,
$G_{\lambda,\pm}^e\sim G_{\lambda,\pm}^o$. Dropping
the even and odd superscripts `o' and `e', we have,
\begin{eqnarray}
G_{\lambda,+} &=& \sum_{m=0}^\infty (-\cos\beta L_m +\sin \beta
K_m)D_m,
\nonumber \\
G_{\lambda,-} &=& \sum_{m=0}^\infty (\sin\beta L_m +\cos \beta
K_m)D_m.
\end{eqnarray}
where
\begin{eqnarray}
&&D_{2k} = \langle 0|\alpha^{\dagger 2k}|0\rangle_\alpha=\langle
0|\overline{\alpha}^{\dagger
2k}|0\rangle_{\overline{\alpha}}=\frac{1}{\sqrt{u}} \frac{(2k)!}{2^k
k!}\frac{v^k}{u^k}, \nonumber \\
&&D_{2k+1} = \langle 1|\alpha^{\dagger 2k}|1\rangle_\alpha=-i\langle
1|\overline{\alpha}^{\dagger
2k}|1\rangle_{\overline{\alpha}}=\frac{1}{u^{3/2}}\frac{(2k+1)!}{2^k
k!}\frac{v^k}{u^{k}}, \nonumber
\end{eqnarray}
where $k=0,1,2,...$.
The results are consistent with   the isotropic two-photon
Rabi model ones, $\lambda=1$\cite{Chen}.

We comment that the construction of $G$-functions as in (\ref{G-functions}) lies ultimately on the fact that the spectral problem for the model (\ref{H2modeW}) can be recast  (through a differential realization of $su(1,1)$) to a differential  equation of the Heun type. The analytical properties of the $G$-functions, giving in turn the eigenvalues of the model, correspond to  specific conditions of analyticity that  Heun functions must fulfill to be  indeed well behaved solution of the  spectral differential equation\cite{Braak}. See \cite{batchelor} for a neat  derivation of the G-functions  as properties of the Heun functions.

\begin{figure}[ht]
  \includegraphics[width=8cm]{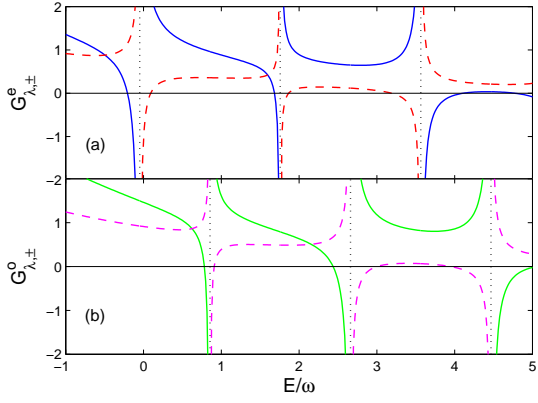}\\
\caption{(color online) Function $G_{\lambda,\pm}^{e,o}$ of
two-photon anisotropic Rabi model for $\omega=1$, $\Delta=0.2$,
$g=0.3$, $\lambda=0.25$ for even (a) odd (b) Fock subspaces,
respectively. The blue (green) lines are $G_+^{e}$ ($G_+^{e}$), and
red (purple) dashed lines are $G_-^{e}$ ($G_-^{o}$), the poles of
$G_{\pm}^{n_0}$ have labeled by dotted lines.}\label{GF2photon}
\end{figure}

As for the single-photon Rabi model\cite{Braak},
the poles of $G(z)$ provide  the eigenvalues of the uncoupled bosonic mode, $\Delta=0$. This is obtained by putting  $a_m =0$
in Eq.(\ref{2photonK_m}):
\begin{eqnarray}
E_{\lambda,m}^{pole}=\omega \eta' (m+\frac{1}{2})-\frac{\omega}{2}.
\end{eqnarray}
where $\eta'=\eta \left[1-(1-\lambda)^2\frac{g^2}{\omega^2}
\right]$, and $m=0,2,4,...$ for $G_{\pm}^e$, $m=1,3,5,...$ for
$G_{\pm}^o$, respectively, as shown in Fig. \ref{GF2photon}.
In the full fledged interacting case, the `regular spectrum'  are given by the zeros of
$G_{\lambda,\pm}^{e,o}$ as $E_{\lambda,\pm}^{e,o}$.  The `irregular spectrum'   providing  the well known  isolated integrability conditions in the parameter space
 $\Delta$, $g$ entails $K_{m+1}(E_{\lambda,m}^{pole})=0$, for  $m=0,1,2,...$. Such points emerge as degeneracy point between odd and even  sectors.

\begin{figure}[ht]
  \includegraphics[width=8cm]{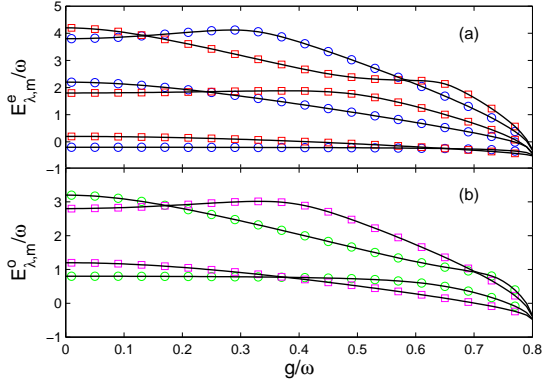}\\
\caption{(color online) Spectrum of the two-photon anisotropic Rabi
model for $\Delta=0.2$, $\lambda=0.25$ in even subspace $m=0,2,4$
(a) and odd subspace $m=1,3$ (b), The solid lines are the numerical
solutions by diagonalizing Hamiltonian, and eigenvalues
$E_{\lambda,+}^{e}$ ($E_{\lambda,+}^{o}$) and $E_{\lambda,-}^{e}$
($E_{\lambda,-}^{o}$) are labeled by blue (green) circle and red
(purple) square, respectively. }\label{energy2photon}
\end{figure}

The first Juddian solution can be obtained from $b_0=0$ and $b_1=0$,
\begin{equation}
\Delta=(1-\lambda^2)\frac{g^2}{2\omega}, \qquad
(1-\lambda^2)\frac{3g^2}{2\omega},
\end{equation}
for $m=0,1$, respectively. For the parameter in Fig.
\ref{energy2photon}, $\Delta=0.2$, $\lambda=0.25$, we get
$g_0=\frac{4\sqrt{2}}{5\sqrt{3}}\approx0.6532$,
$E_0^{pole}=\frac{1}{2}\sqrt{\frac{19}{75}}-\frac{1}{2}\approx
-0.2483$, and $g_1=\frac{4\sqrt{2}}{15}\approx 0.3771$,
$E_1^{pole}=\frac{3}{2}\sqrt{\frac{161}{225}}-\frac{1}{2}\approx
0.7689$. The positions of level-crossing are consistent with them
very well.

\section{Criticality in the  energy levels}
Despite our  model describes a coupling between a bosonic  field and a two levels system in `zero dimension', 
the energy levels of the system display a 'critical'  behavior resembling very much that one occurring  in quantum  phase transitions of  many-body systems. 

\subsection{Discontinuity of  entanglement entropy}
First,  we discuss the wave function:
\begin{eqnarray} \label{PsiC}
\Psi_C=\left(\begin{array}{c} C\varphi_1 \\ \overline{\varphi}_1
\end{array} \right), \ {\rm or} ~~~~  \overline{\Psi}_C =\left(\begin{array}{c} \overline{\varphi}_2 \\ C\varphi_2
\end{array} \right),
\end{eqnarray}
where the constant $C=\pm 1$ or $C=\pm i$.
%
%
%
%
In each parity sector, $\Psi_C$ reads as in (\ref{psi-c}).
At parity degenerate point, the wave function with fixed parity $C$
can be determined as follows,
\begin{eqnarray}
\Psi_C=\left(\begin{array}{c} \sum_{m=0}^M C L_m|m\rangle\rangle +K_m |\overline{m}\rangle\rangle \\
\sum_{m=0}^M C K_m|m\rangle\rangle +L_m |\overline{m}\rangle\rangle
\end{array} \right)
\end{eqnarray}
where $M$ is the truncated number, $|m\rangle\rangle$ denotes $|m\rangle_\alpha$,
in the appendix it can denote $|n_0+m\rangle_b$ in two-mode case, such
that the wave functions can be written in a unified form,
similarly $|\overline{m}\rangle\rangle$ is the corresponding basis under
transformation $a^\dagger \rightarrow i a^\dagger$.
%
%
For our anisotropic model, the first
energy crossing point corresponds to  $K_{m+1}(E_m^{pole})=0$, $m=0,1$:
\begin{eqnarray}
K_0 =1, \qquad L_0=\tan 2\beta,
\end{eqnarray}
and the corresponding wave function is
\begin{eqnarray}
\Phi=\left(\begin{array}{c} \tan 2\beta |0\rangle \rangle \\
|0\rangle \rangle \end{array} \right) .
\end{eqnarray}
Therefore, the wave function of the $Z_2$ Hamiltonian is
\begin{eqnarray}
\Psi=V^\dagger \Phi \propto \left(\begin{array}{c} (\cos\beta+\sin\beta)|0\rangle \rangle \\
(\cos\beta-\sin\beta)|0\rangle \rangle
\end{array} \right),
\end{eqnarray}
another eigenfunction is
\begin{eqnarray}
\overline{\Psi}\propto \left(\begin{array}{c}
(\cos\beta-\sin\beta)|\overline{0}\rangle \rangle \\
(\cos\beta+\sin\beta)|\overline{0}\rangle \rangle
\end{array} \right)
\end{eqnarray}
The eigenfunction labeled by parity $C$ can be written as,
\begin{equation}
\Psi_C=C\Psi +\overline{\Psi}.
\end{equation}
\begin{figure}[ht]
  \includegraphics[width=8cm]{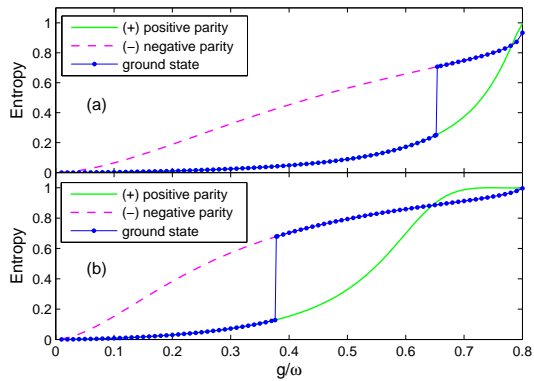}\\
\caption{ Entanglement entropy of the ground and first excited state
of 2-photon anisotropic Rabi model of even (a) and odd (b) Fock
subspace of Bogolubov operator, respectively, the parameter is the
same as Fig. \ref{energy2photon}. According a level crossing occurs
at $g=\frac{4\sqrt{2}}{5\sqrt{3}}$ (a) and $\frac{4\sqrt{2}}{15}$
(b) in the energy level, there is a sharp discontinuous parity
change of the ground state from $+$ to $-$ as $g$ become larger, as
shown by a blue-dot line. }\label{en2photon_lam}
\end{figure}

The entanglement entropy is
\begin{equation}
S=-{\rm tr} \rho \log \rho ,
\end{equation}
where $\rho $ is the reduced density matrix obtained by tracing out the bosonic degrees of freedom $\rho={\rm tr}_{bosonic}(|\Psi\rangle \langle \Psi|)$ \cite{Amico-RMP}.
For the first two levels:  $S=-\sum_{i=1}^2
\lambda_i\log_2(\lambda_i)$, where $\lambda_i$ is
the eigenvalue of $\rho$. See Fig. \ref{en2photon_lam}, we can
find that the entanglement entropy of the ground state jumps at
a critical point. The size of the jump depends on
anisotropic parameter $\lambda$.

We comment that the  discontinuity displayed by the entanglement occurs with the same mechanism (level crossing) in which first order phase transitions occur in many body systems \cite{Amico-RMP,Amico-Nature,JianCui,Franchini}. We also comment that  the parity of the
ground state changes at this critical point.

\subsection{Energy levels condensation}

With a logic that is similar to the one applied to second order quantum phase transitions, we analyze the behavior of the energy spectrum of $H$ for  spin-boson coupling in the neighborhood of   $|g|/\omega=g_c=\frac{1}{|1+\lambda|}$ (with no lack of generality,  we set  $g>0$,
$\lambda>0$, and $\omega=1$).
Indeed, by inspection of Eq.(\ref{z}), we notice that $\eta=0$ at $g_c$. This implies that  all the levels condense at    $E_0=-\frac{1}{2}$, un-respective  of  $m$, therefore  with  a clear parity symmetry breaking. The transcendental function $G_{\lambda,\pm}^{e,o}$,  in turn, displays a pole of higher order.

This scenario indicates that  $\langle a \rangle$ can be non vanishing at $g=g_c$ meaning that the bosonic mode is macroscopically occupied. Therefore, such energy condensation implies  super-radiance.

\begin{figure}
  \includegraphics[width=8cm]{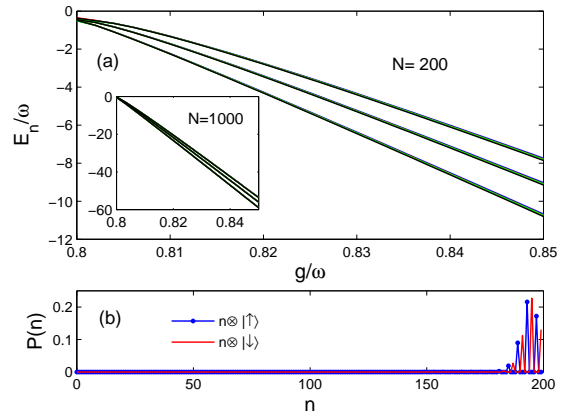}\\
\caption{(color online)
 (a) Energy levels for $g>g_c$ and large $n$.
The lowest 12 energy levels  for $g>g_c=0.8$ are displayed which
seem be grouped into three separate bold curves,
each bold curve actually includes four lines colored
black, green, red and blue, from down to up, corresponding to the
subspace of $p= -i$, $-1$, $i$ and $1$, respectively. The inset displays the results for  $N=1000$.
(b) Photon number probability
of the ground state for $g=0.85$ and $N=200$. }\label{after_gc}
\end{figure}

Notwithstanding we cannot analyze the energy for $g>g_c$ with our exact solution (because of the topological constraint coming from the non-compactness of $su(1,1)$ algebra discussed in Sec(\ref{sect2})),
 the numerical data shown in Fig. \ref{after_gc},  indicate that  the low energy levels
decrease linearly, with a slope depending  on   the total bosonic  number as $-N^{x}$ with $x\sim 1.1$. If $N\rightarrow \infty$,
the curve will be a vertical line.
%
Some insight in the limit of large $n$ can be acquired by resorting  to the RWA.Accordingly, the  lowest energy is
$E_0 = (n+1)\omega/2 - \sqrt{(\omega -\Delta)^2 +g^2(n+1)(n+2)} \approx (n+1)\omega/2 -g(n+\frac{3}{2}) \approx n(g_c -g)$,
in the limit of $n\gg 1$ (the term of $\omega - \Delta$ can be neglected in the resonant case).
This indicates that for $g>g_c$ the ground state energy  decreases as $E_0\approx -n\omega $ for large $n$ (Fig. \ref{after_gc}(a)).

 In the Fig.\ref{after_gc}(b) where $g=0.85$ which is larger than $g_c$,
the photon number probability of ground state appears to be
equally distributed among a large  number of modes $n$. Accordingly,
the entanglement entropy saturates the bound 1, as shown in Fig. \ref{en2photon_lam}.

We remark that such a parity change occurs also in the anisotropic Rabi model proposed in Ref.\cite{Xie}.

Similar findings were recently reported for  the isotropic Rabi model by Plenio and coworkers\cite{qphasetransition}.

\section{Discussions and summary}
In this paper we studied a two-boson spin-boson model of the Rabi type in which the rotating and counter-rotating  spin-boson couplings act with two different parameters.
The exact solution of the model is provided through a variation of the Braak method. We remark that our formalism can be straightforwardly applied to  two-modes Rabi models enjoying the same algebraic structure $su(1,1)\otimes Z_2$ of the two-boson case (see the appendix). The case with explicit parity symmetry breaking (inclusion of a term proportional to $\sigma_x$) could be approached as well within our scheme with minor changes\cite{Xie}. 

We comment how  the spectrum of the model is found modulo  certain restriction in the system parameters (Eq.(\ref{condition})). Ultimately, such a constraint arises because the procedure leading to the exact solution involves a diagonalization in $su(1,1)$  which is a non-compact Lie algebra. In order to go beyond such restriction, one should use the continuous representation of Limblad-Nagel  instead of the discrete one\cite{Limblad-Nagel}. To this route, a separate study will be dedicated.

Remarkably, despite the Hamiltonian describes just two interacting degrees of freedom, the energy spectrum of the system displays  characteristic quantum critical features.  
We believe that this is possible because of the specific $Z_2$ symmetry of the system,  acting in a `zero dimension' space. Specifically, a 'first order quantum phase transition' occurs  because two states belonging to different parity sectors with the lowest energies cross. Such a level crossing is visible as a discontinuity of the entanglement entropy (obtained tracing out the bosonic degree of freedom).  The other  transition occurs at higher values of the spin-boson coupling strength. It is displayed as a parity symmetry breaking transition allowing a macroscopic occupation of the bossing mode with a `off diagonal order' $\langle a \rangle\neq 0$.
By exact analysis of the entanglement entropy we established that the energy condesation occurs continuously,
sharing similarities with the super-radiance phenomenon.
Finding the critical properties of the system within the exact formalism defines a major direction for future investigation.

%
\section*{Acknowledgement}

We thank discussions with Qiong-Tao Xie. This work is supported
by MOST of China (Grant Nos. 2016YFA0302104 and
2016YFA0300600);  NSFC (Grant Nos.91536108); Chinese Academy of Sciences (Grant XDB01010000 and XDB21030300); Ministry of Education, Singapore Academic
Research  Fund  Tier  2  (Grant  No.   MOE2015-T2-1-101) and under the Research Centres of Excellence programme.

\begin{appendix}

\section{A realization of the anisotropic two-photon Rabi model in circuit QED.}
In this section, we provide details on how a two-photon Hamiltonian can be realized through  circuit QED.

We refer to the  phase qubit, interacting with a microwave field\cite{phaseQubit,Chuang}.
For such system, three or more levels participate to the quantum dynamics. 
Phase qubit  can be realized with  a Josephson junction in a high-inductance superconducting loop biased with a flux sufficiently large that the phase across the junction sees a potential analogous to that found for the current-biased junction.
The energy differences between the ground and first excited states $\omega_{01}$,  and
the first and the second excitation  $\omega_{12} $ are different each other. By tuning  the 
boson field frequency to  half of the difference between the ground and second excitation energy,  the first excitation is excluded from the dynamics.   The transition between the  ground and the second excitations can occur by  two photons.
In this regime, the Hamiltonian of the magnetic flux-biased phase qubit can be written as
\begin{eqnarray}\label{H2photon0728}
H_{ph}=\omega_r a^\dagger a +\omega_q \sigma^+ \sigma^- +g(\cos\theta
\sigma_z -\sin\theta \sigma_x)(a^2 +a^{\dagger 2}), \nonumber
\end{eqnarray}
Experiments, nowadays can access the  ultra-strong coupling regime.   
Indeed, typical  numbers involved  in the experiments are 
$\omega_{01}/2\pi=5.5$GHz, $\omega_{12}=4.5$GHz, and coupling
constant $g=0.1$GHz. These parameters  implies a detectable Bloch-Siegert shift  $\sim g^2/\Delta=1 MHz$.

Anisotropies in the spin-boson coupling  can arise from mutual inductance between the phase qubit and the controlling SQUID. 
In this case the Hamiltonian of the circuit is $H_{ph}+M (a^2 -a^{\dagger 2})$, which, apart from the $g\cos\theta
\sigma_z (a^2 +a^{\dagger 2})$, term can be recast to the  anisotropic two-boson Rabi model Eq.(\ref{H2photon}). 

%

\bigskip

\section{The  anisotropic 2-mode Rabi model}

The Hamiltonian for the two-mode Rabi model reads

\begin{eqnarray}
\label{H2mode}
H_{2-m}&=& a_1^\dagger a_1 + a_2^\dagger a_2 +\Delta
\sigma_z +g(\sigma^+ a_1 a_2 +\sigma^- a_1^{\dagger} a_2^\dagger) \nonumber \\
&& \qquad \qquad +g\lambda (\sigma^+ a_1^{\dagger} a_2^\dagger +
\sigma^- a_1 a_2).
\end{eqnarray}
With two modes in this Hamiltonian, the $su(1,1)$ Lie algebra is spanned by
\begin{equation}
K_0=\frac{1}{2}(a_1^\dagger a_1 +a_2^\dagger a_2 +1), \
K_+=a_1^\dagger a_2^\dagger, \  K_-=a_1 a_2 \;. \nonumber
\end{equation}
Here, $\kappa=\frac{1}{2},1,\frac{3}{2}, \ldots$,
$\kappa>0$ for two-mode case, every $\kappa$ corresponds an
irreducible subspace. These subspace can also be written as
$\{|n_0+n, n\rangle \}$, $n_0,n=0,1,2,...$, here $|n_1,n_2\rangle$
($n_1 \geq n_2$) is the Fock state of the boson operator $a_1$,
$a_2$. We note that $n_1 \leq n_2$ give a set of equivalent subspace,
$\{|n, n_0+n\rangle \}$,  giving  degenerate eigenvalues and
eigenfunctions; in the following we will work with the first choice. Thus
we can also recognize $n_0$ as the subspace index, the relation of
$n_0$ and $\kappa$ is $n_0=2\kappa -1$.

The Hamiltonian enjoys the $Z_2$ symmetry generated by
\begin{eqnarray}
\label{twomode_Z2}
\Pi_{2-m}=e^{i\pi[a_2^\dagger a_2
+\frac{1}{2}(\sigma_z+1)]}=-\sigma_z \cos (\pi a_2^\dagger a_2)
\end{eqnarray}
which have value $p=\pm 1$ in the corresponding subspaces of $\kappa $.

\subsection{Exact solution of the anisotropic 2-mode  Rabi model}

The exact analysis of the two mode Rabi model (\ref{H2mode}) proceed along very similar lines we followed for the two-boson case.
We apply  $U$ and the Bogolubov
transformation:
\begin{widetext}
\begin{eqnarray} \label{2mode_lamH2}
U^\dagger (H) U= \left(\begin{array}{cc} \omega a_1^\dagger a_1
+\omega a_2^\dagger a_2 +p +r(a_1 a_2 +a_1^{\dagger} a_2^\dagger) &
-q +s a_1 a_2 +t a_1^{\dagger} a_2^\dagger
\\ -q +s
a_1^{\dagger} a_2^\dagger +t a_1 a_2 & \omega a_1^\dagger a_1
+\omega a_2^\dagger a_2 -p -r(a_1 a_2 +a_1^{\dagger} a_2^\dagger)
\end{array} \right)
\end{eqnarray}
\end{widetext}
\begin{eqnarray}
b_1=ua_1+va_2^\dagger, \qquad b_2=ua_2 +va_1^\dagger,
\end{eqnarray}
with
\begin{eqnarray}
|g|< \frac{2\omega}{|1+\lambda|},
\end{eqnarray}
we can get
\begin{eqnarray}
u=\sqrt{\frac{1+\zeta}{2\zeta}}, \ v=\sqrt{\frac{1-\zeta}{2\zeta}},
\ \cos 2\beta=\frac{1-\lambda}{1+\lambda}\zeta,
\end{eqnarray}
where
\begin{eqnarray}
\zeta=\sqrt{\frac{1-(1+\lambda)^2 g^2/4\omega^2}{1-(1-\lambda)^2
g^2/4\omega^2}},
\end{eqnarray}


The
vacuum state can be found in $b_1 b_2|n_0,0\rangle_b=0$, as
\begin{eqnarray}
|n_0,0\rangle_b=\frac{1}{u^{n_0+1}}\sum_{n=0}^\infty \left(
-\frac{v}{u} \right)^n \sqrt{\frac{(n_0+n)!}{n_0!
n!}}|n_0+n,n\rangle,
\end{eqnarray}
which also have property $b_1
|n_0,0\rangle_b=\sqrt{n_0}|n_0-1,0\rangle_b$. Here, $1/u^{n_0+1}$ is
the normalized coefficient, which can be found with the Taylor
expansion,
\begin{eqnarray}
\frac{1}{(1-x)^{n_0+1}}=1 +(n_0+1)x +\frac{(n_0+2)(n_0+1)}{2}x
+\cdots, \nonumber
\end{eqnarray}
where $x=v^2/u^2$. Actually, $|n_0,0\rangle_b$ is a two-mode
squeezed vacuum state, which can be generated by the 2-mode squeeze
operator $S_2(\xi)$ on the vacuum
\begin{eqnarray}
|n_0,0\rangle_b =S_2(\xi)|n_0,0\rangle =\exp\left(\xi a_1^\dagger
a_2^\dagger -\xi^* a_1 a_2 \right)|n_0,0\rangle, \nonumber
\end{eqnarray}
where $u=e^{|\xi|^2/2}$ and $v=-\xi e^{-|\xi|^2/2}$. In quantum
optics $S_2(\xi)$ is associated with degenerate parametric
amplification \cite{Scully}.

The Fock states can be constructed as
\begin{eqnarray}
|n_0+m,m\rangle_b=(b_1^\dagger b_2^\dagger)^m |n_0,0\rangle_b,
\end{eqnarray}
Note that, the normalized coefficient $\sqrt{\frac{n_0 !}{(n_0+m)!
m!}}$ is eliminated for the simplicity.

The ansatz for the eigenstates of $(H)$ is
\begin{eqnarray}\label{2mode_lamphi}
\left(\begin{array}{c} \phi_1 \\ \phi_2 \end{array} \right)
=\left(\begin{array}{c} \sum_{m=0}^\infty L_m |n_0+m,m\rangle_b
\\ \sum_{m=0}^\infty K_m |n_0+m,m\rangle_b \end{array}
\right)
\end{eqnarray}

Because of the $Z_2$ symmetry, the constant $C$ has only two values
$\pm 1$. So we can construct the transcendental function as:
\begin{eqnarray}
G_{\lambda,\pm}^{n_0} =\langle n_0, 0|[\overline{\varphi}_2 \mp
\varphi_1 ],
\end{eqnarray}
use the coefficients above we can obtain G-function as:
\begin{eqnarray}
G_{\lambda,+}^{n_0} &=& \sum_{m=0}^\infty (-\cos\beta L_m +\sin
\beta K_m)D_m,
\nonumber \\
G_{\lambda,-}^{n_0} &=& \sum_{m=0}^\infty (\sin\beta L_m +\cos \beta
K_m)D_m.
\end{eqnarray}
where the coefficients $D_m$ is as follows
\begin{eqnarray}\label{Dm2mode}
D_m &=& \langle n_0, 0|(b_1^\dagger
b_2^\dagger)^m|n_0,0\rangle_b=\langle n_0, 0|(\overline{b}_1^\dagger
\overline{b}_2^\dagger)^m|n_0,0\rangle_{\overline{b}} \nonumber \\
&=& \frac{1}{u^{n_0+1}} \left(\frac{v}{u}\right)^m
\frac{(n_0+m)!}{n_0!}.
\end{eqnarray}

\begin{figure}[ht]
  \includegraphics[width=8cm]{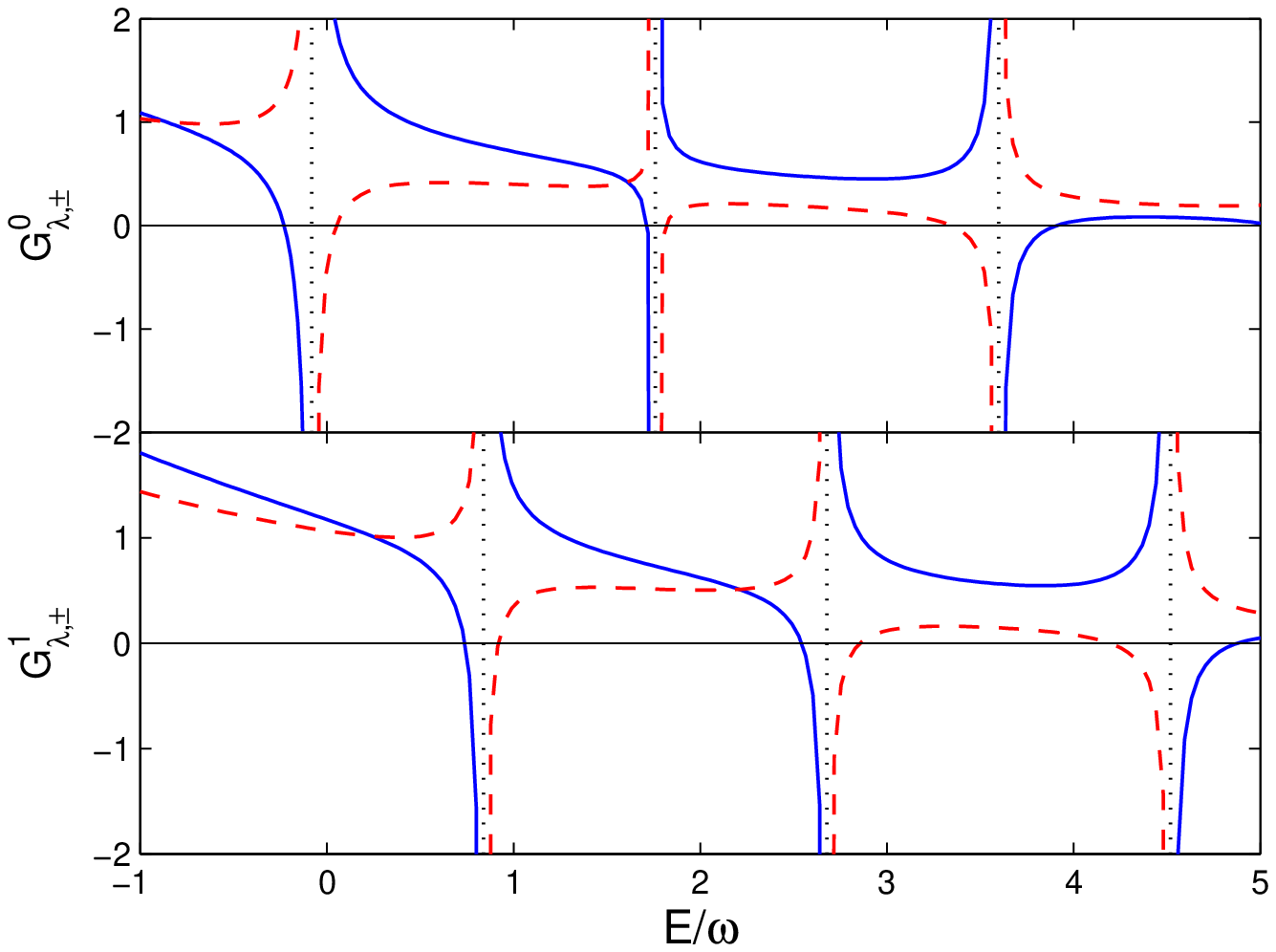}\\
\caption{(color online) Function $G_{\lambda,\pm}^{n_0}$ of two-mode
anisotropic Rabi model for $\omega=1$, $\Delta=0.2$, $g=0.5$,
$\lambda=0.5$ for $n_0=0$ (a) $n_0=1$ (b), respectively. The blue
lines are $G_+^{n_0}$, and red dashed lines are $G_-^{n_0}$, the
poles of $G_{\pm}^{n_0}$ have labeled by dotted
lines.}\label{GF2mode_lam}
\end{figure}

The analytical property of $G_{\lambda,\pm}^{n_0}$ is similar to it
in the isotropic case, as shown in Fig. \ref{GF2mode_lam}. The poles
of $G_{\lambda,\pm}^{n_0}$ can be found similar as that for two-photon case,
\begin{eqnarray}
E_{\lambda,m}^{pole}=\omega \zeta' (n_0+2m+1)-\omega.
\end{eqnarray}
where $\zeta'=\zeta \left[1-(1-\lambda)^2\frac{g^2}{4\omega^2}
\right]$. The energy spectra has divided into two parts, one is the
regular case, which correspond to the zeros of
$G_{\lambda,\pm}^{n_0}$ as $E_{\lambda,m,\pm}^{n_0}$. The other is
the irregular case, which correspond to the level-crossing points,
i.e., Juddian solutions. They can be found similarly as $K_{m+1}
(E_{\lambda,m}^{pole})=0$, then the numerator and denominator of
$G_{\pm}^{n_0}$ are both 0. However, there is a few difference, in
anisotropic case the first degenerate point is at $m=0$, not $m=1$
in isotropic case, because of the parameter $\lambda$ of the
counter-rotating term.

\begin{figure}[ht]
  \includegraphics[width=8cm]{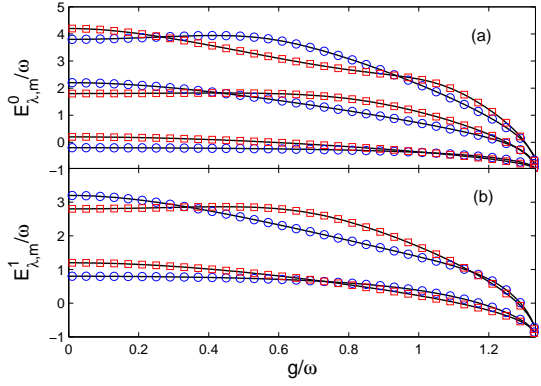}\\
\caption{(color online) $E_{\lambda,m}^{n_0}$ vs. $g$ of $n_0=0$,
$m=0,1,2$ (a) and $n_0=1$, $m=0,1$ (b) for $\omega=1$, $\Delta=0.2$,
$\lambda=0.5$. The solid lines are the numerical solution, and
eigenvalues of parity $+$ and $-$ are labeled by blue circle and red
square, respectively. }\label{energy2mode_lam}
\end{figure}

The Juddian solution at $m=0$ can be found as
\begin{eqnarray}
E_{\lambda,0}^{n_0}=\omega \eta' (n_0+1) -\omega,
\end{eqnarray}
if $\Delta$ is known, from $b_0=0$ the especial value of $g$ can be
found as
\begin{eqnarray}
|g|=\sqrt{\frac{4\Delta \omega}{(n_0+1)(1-\lambda^2)}},
\end{eqnarray}

As shown in Fig. \ref{energy2mode_lam}, we can exactly determine the
first level-crossing points at subspace $n_0=0,1$ for parameter
$\Delta=0.2$, $\lambda=0.5$. In $n_0=0$ subspace, it is
$g=\frac{4}{\sqrt{15}}\approx 1.0328$,
$E=\sqrt{\frac{28}{75}}-1\approx -0.3890$. And
$g=\frac{2\sqrt{2}}{\sqrt{15}}\approx 0.7303$,
$E=\sqrt{\frac{203}{75}}-1 \approx 0.6452$ for $n_0=1$ case. The
wavefunctions at these degenerate points will be discussed in detail
later.

\subsection{Entanglement  entropy}

For example, in isotropic 2-mode and 2-photon case, the first level
crossing occurs in $m=1$,
\begin{eqnarray}
K_0=1, \qquad L_0=-\frac{\Delta}{2\omega\eta}.
\end{eqnarray}
So the eigenfunction can be written as
\begin{eqnarray}
\Psi=\left(\begin{array}{c} L_0 |0\rangle \rangle \\ K_0 |0\rangle
\rangle  \end{array}\right), \qquad or \qquad \left(\begin{array}{c}
K_0 |\overline{0}\rangle \rangle \\ L_0 |\overline{0} \rangle
\rangle
\end{array} \right),
\end{eqnarray}
where basis $|0\rangle \rangle$ can be $|n_0,0\rangle_b$,
$|0\rangle_\alpha$ and $|1\rangle_\alpha$, and
$|\overline{0}\rangle\rangle$ is corresponding case of
$|0\rangle\rangle$.

Using parity they can be written as
\begin{eqnarray}
\Psi_C=\left(\begin{array}{c} C L_0 |0\rangle \rangle+K_0
|\overline{0}\rangle \rangle \\ C K_0 |0\rangle \rangle+L_0
|\overline{0}\rangle \rangle \end{array}  \right),
\label{psi-c}
\end{eqnarray}
where $C=\pm 1, \pm i$, the two eigenfunctions are orthogonal to
each other.

\begin{figure}[ht]
  \includegraphics[width=8cm]{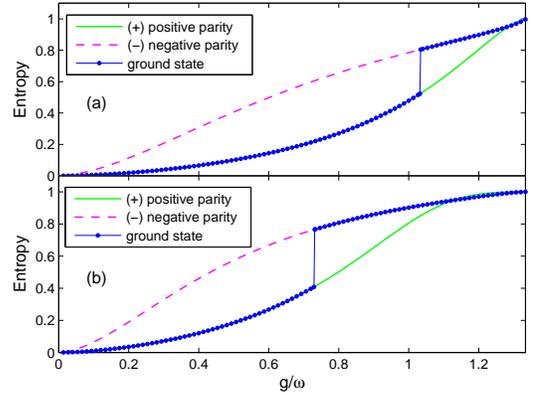}\\
\caption{ Entanglement entropy of the ground and first excited state
of 2-mode anisotropic Rabi model of $n_0=0$ (a) and $n_0=1$ (b),
respectively, the parameter is the same as Fig.
\ref{energy2mode_lam}. The level crossings occur at
$g=\frac{4}{\sqrt{15}}$ (a) and $\frac{2\sqrt{2}}{\sqrt{15}}$ (b) in
the energy level, there is a sharp discontinuous parity change of
the ground state from $+$ to $-$ as $g$ become larger, as shown by a
blue-dot line. }\label{en2mode_lam}
\end{figure}

We can study the entanglement entropy of the ground state.
Similar as in two-photon case, here we can also find that the
entanglement entropy of the ground state jumps at a critical point,
see Fig. (\ref{en2mode_lam}).

\end{appendix}

\end{document}